\documentclass[twocolumn,twoside,9pt]{nih}	
\newcommand{\manuscript}[1]{}	

\usepackage{times}
\usepackage{natbib}
\usepackage{amssymb,amsfonts,amsmath,amscd}
\usepackage{amsthm}
\usepackage{graphicx}
\usepackage{url}
\usepackage{hyperref}
\usepackage[usenames]{color}
\usepackage{mathtools} 
\usepackage{mathrsfs}		 

\newcommand{\ERS}{{\sf ERS}}	
\newcommand{\LCI}{{\sf LCI}}	
\newcommand{\LSCD}{{\sf{LSCD}}}	
\newcommand{\NI}{{\sf{NI}}}	
\newcommand{\RR}{{\sf{RR}}}	

\newcommand{\TV}{{Tomasetti:and:Vogelstein:2015:Variation}}
\newcommand{\Xf}{{\sf X}}

\newcommand {\dsty}{\displaystyle}

\newcommand{\Cov}{\mbox{\rm Cov}}

\newcommand{\Nc}{{\cal N}}

\newcommand{\Uc}{{\cal U}}

\newcommand{\Var}{{\sf Var}}

\newcommand{\an}[1]{\begin{align}#1\end{align}}
\newcommand{\ab}[1]{\begin{align*}#1\end{align*}}

\newcommand{\Quote}[1]{\begin{quote}#1\end{quote}}

\newcommand{\dspfrac}[2]{\frac{\displaystyle #1}{\displaystyle #2} }

\newcommand{\enumlist}[1]{\begin{enumerate}#1\end{enumerate}}

\newcommand{\eqdef}{:=}

\newcommand{\oldversion}[1]{}

\newcommand{\hide}[1]{}	
\newcommand{\hidex}[1]{\ }	
\newcommand{\temphide}[1]{}	

\newcommand{\kwsp}{; }

\newcommand{\urlsmall}[1]{{\small\url{#1}}}

\newfont{\gilfont}{cmsy10 scaled\magstep0}

\newcommand{\Psf}{\mathsf{P}}

\newcommand{\desclistZero}[1]{\begin{description}\setlength{\itemsep}{-.0mm}#1\end{description}}

\hide{ 
\renewcommand{\citet}{\cite}
\renewcommand{\citep}{\cite}

\renewcommand{\citeauthor}{\cite}

}

\newtheorem{OpenQuestion}{OpenQuestion}

\newtheorem{Theorem:}[Theorem]{Theorem:}
\newtheorem{Conjecture:}[Theorem]{Conjecture:}
\newtheorem{Corollary:}[Theorem]{Corollary:}
\newtheorem{Definition:}[Theorem]{Definition:}
\newtheorem{Lemma:}[Theorem]{Lemma:}
\newtheorem{Paradox:}[Theorem]{Paradox:}
\newtheorem{Principle:}[Theorem]{Principle:}
\newtheorem{Proposition:}[Theorem]{Proposition:}
\newtheorem{Recursion:}[Theorem]{Recursion:}
\newtheorem{Result:}[Theorem]{Result:}

\newtheorem{Theorem-InOrder}[Theorem]{Theorem}
\newtheorem{Conjecture-InOrder}[Theorem]{Conjecture}
\newtheorem{Corollary-InOrder}[Theorem]{Corollary}
\newtheorem{Definition-InOrder}[Theorem]{Definition}
\newtheorem{Lemma-InOrder}[Theorem]{Lemma}
\newtheorem{Paradox-InOrder}[Theorem]{Paradox}
\newtheorem{Principle-InOrder}[Theorem]{Principle}
\newtheorem{Proposition-InOrder}[Theorem]{Proposition}
\newtheorem{Recursion-InOrder}[Theorem]{Recursion}
\newtheorem{Result-InOrder}[Theorem]{Result}
\newtheorem{Remark-InOrder}[Theorem]{Remark}
\newtheorem{Counterexample-InOrder} [Theorem]{Counterexample}
\newtheorem{Goal-InOrder} [Theorem]{Goal}

\hide{
\newtheorem*{Theorem*}{Theorem}
\newtheorem*{Conjecture*}{Conjecture}
\newtheorem*{Corollary*}{Corollary}
\newtheorem*{Proposition*}{Proposition}
\newtheorem*{Definition*}{Definition}
\newtheorem*{Lemma*}{Lemma}
\newtheorem*{Paradox*}{Paradox}
\newtheorem*{Principle*}{Principle}
\newtheorem*{Recursion*}{Recursion}
\newtheorem*{Result*}{Result}
\newtheorem*{BlankTheorem*}{}

}

\theoremstyle{plain}

\theoremstyle{plain}

\begin{document}
\manuscript{\begin{spacing}{1.9}}	
\title{Statistical Problems in a Paper on \\
Variation In Cancer Risk Among Tissues,\\
and New Discoveries 
}
\author{{\normalsize\sc Lee Altenberg} \\
{\normalsize The KLI Institute, \href{mailto:Lee.Altenberg@kli.ac.at}{Lee.Altenberg@kli.ac.at}. }}
\maketitle
\sloppy
\sloppypar

\begin{abstract}\rm
\citet{\TV} collected data on 31 different tissue types and found a correlation of $0.8$ between the logarithms of the incidence of cancer (\LCI), and the estimated number of stem cell divisions in those tissues (\LSCD).  Some of their conclusions however are statistically erroneous.  Their excess risk score, ``\ERS'' ($\log_{10} \LCI \times \log_{10} \LSCD$), is non-monotonic under a change of time units for the rates, which renders meaningless the results derived from it, including a cluster of 22 ``R-tumor'' types for which they conclude, ``primary prevention measures are not likely to be very effective''.  Further, $r=0.8$ is consistent with the three orders of magnitude variation in other unmeasured factors, leaving room for the possibility of primary prevention if such factors can be intervened upon.  Further exploration of the data reveals additional findings:  (1) that $\LCI$ grows at approximately the square root of $\LSCD$, which may provide a clue to the biology; (2) among different possible combinations of the primary data, the one maximizing the correlations with $\LCI$ is almost precisely the formula used by \citeauthor{\TV} to estimate $\LSCD$, giving support to stem cell divisions as an independent factor in carcinogenesis, while not excluding other such factors.

\renewcommand{\kwsp}{,\ }
\end{abstract}


\section*{Introduction}
A paper recently published in \emph{Science}, ``Variation in cancer risk among tissues can be explained by the number of stem cell divisions'' by  \citet{Tomasetti:and:Vogelstein:2015:Variation}, investigates the relationship between the number of stem cell divisions in various tissue types and the incidence of tumors in them.  They find a strong correlation, $0.8$, between the logarithms of these two values (both Spearman's rank correlation $\rho$ and Pearson's linear correlation $r$).  They use the product of these logarithms, which they call the ``extra risk score'' (\ERS) to classify tumor types into ``R-tumors'', where ``stochastic factors, presumably related to errors during DNA replication, most strongly appear to affect their risk,'' and ``D-tumors'', where ``deterministic factors such as environmental mutagens or hereditary predispositions strongly affect their risk''.  On this basis, they give a prognosis for the likely success of interventions to prevent these tumor types:

\Quote{These results could have important
public health implications. One of the
most promising avenues for reducing cancer
deaths is through prevention. How successful can
such approaches be? The maximum fraction of
tumors that are preventable through primary
prevention (such as vaccines against infectious
agents or altered lifestyles) may be evaluated
from their ERS. For nonhereditary D-tumors,
this fraction is high and primary prevention could
make a major impact (31). Secondary prevention,
obtainable in principle through early detection,
could further reduce nonhereditary D-tumor-related deaths and is also instrumental for reducing hereditary D-tumor-related deaths. For R-tumors, primary prevention measures are not
likely to be very effective, and secondary prevention
should be the major focus.
}

Here I describe statistical problems with the paper that undermine these conclusions.  These problems are so basic that they ought to have been caught in review, but apparently were not.  

First, any statistic on ``extra risk'' should be invariant under a change in the units of time used to measure the rates.  Their ERS statistic not only fails to be invariant, but is non-monotonic under a simple change of time units in the data.  This renders meaningless all the conclusions based on it.  

Second, there is a fundamental misunderstanding of what high correlations imply.  The argument that ``primary prevention measures are not likely to be very effective'' rests on the idea that high correlations between a variable not subject to intervention (number of stem cell divisions) and a target variable (cancer incidence) means that the target variable is mostly nonsusceptible to intervention.  

This is wrong on two counts:  first, because the correlation is between logarithms, it is possible for a second, unmeasured factor to vary, in this case over four orders of magnitude, and still maintain the correlation of $0.8$ for the data.  Second, correlations only put limits on the existing variation of unknown factors;  they have nothing to say about novel interventions that may be developed which change that variation.  This latter point is well developed in \citet{Feldman:and:Lewontin:1975}, an article prompted by the misuse of heritability measures of human intelligence.  The potential misuse of correlation measures in decisions about cancer research prompts the work here.

Putting aside the statistical problems in \citet{Tomasetti:and:Vogelstein:2015:Variation}, further exploration of their data reveal some tantalizing clues to the biology of cancer.  

First, the best fit to the data indicates that cancer incidence grows not in proportion to the number of stem cell divisions in a tissue, but in proportion close to the square root.   Second, an exploration of different combinations of primary data, $s$ and $d$, they use to estimate the lifetime number of stem cell divisions shows the correlations with the lifetime cancer incidence are maximized almost \emph{precisely} by the formula they use, $\LSCD=s(2+d)-2$.  This suggests that their estimate for the number of stem cell divisions or a closely related formula is a real biological factor in the incidence of cancer, while not ruling out the possibility of other central factors.

\section*{The Ill-Behaved ``Extra Risk Score" (ERS)}
\citet{Tomasetti:and:Vogelstein:2015:Variation} introduce their \ERS\ statistic:
\Quote{We next attempted to distinguish the effects of
this stochastic, replicative component from other
causative factors---that is, those due to the external
environment and inherited mutations. For
this purpose, we defined an Òextra risk scoreÓ
(ERS) as the product of the lifetime risk and the
total number of stem cell divisions ($\log_{10}$ values). \ldots

The ERS provides a test of the approach described
in this work. If the ERS for a tissue type is
high---that is, if there is a high cancer risk of that
tissue type relative to its number of stem cell
divisions---then one would expect that environmental
or inherited factors would play a relatively
more important role in that cancerÕs risk
(see the supplementary materials for a detailed
explanation). It was therefore notable that the
tumors with relatively high ERS were those with
known links to specific environmental or hereditary
risk factors (Fig. 2, blue cluster).}

The most straightforward statistic to measure the ``cancer risk of that tissue type relative to its number of stem cell divisions'' would be the ratio of risk to the number of cell divisions, $\LCI/\LSCD$.   On the logarithmic scale this would be $\log_{10} \LCI - \log_{10} \LSCD$.  For unknown reasons, the authors instead devise a statistic $\ERS(\LCI,\LSCD) \eqdef \log_{10} \LCI \times \log_{10} \LSCD$.  They also  considered using $\log_{10} \LCI / \log_{10} \LSCD$ (see their Supplement) but reject it, not on first principles, but because it is ``suboptimal'':
\Quote{ ``Note that using the ratio between the $\log_{10}$ values of $r$ [\LCI] and ${\it lscd}$, instead of the product, would be sub-optimal to estimate the extra risk. \ldots When ERS is defined as the product rather than the ratio, the expected relationship is evident''.}

One expected relationship that should be evident for a measurement of extra risk is that it be invariant under a change of time units.  If the time units were changed from \emph{per lifetime} to \emph{per lifetime}$/ T$, this should be irrelevant to a measure of ``cancer risk of that tissue type relative to its number of stem cell divisions''.

However, we find that \ERS\ is extremely ill-behaved in this regard: it is not invariant, and even worse, it is non-monotonic.  This can be seen from its expansion:
\ab{
\ERS&(\LSCD/T, \ \LCI/T) \\&
= \ERS(\LSCD, \ \LCI) \\&
- T (\log_{10}\LSCD + \log_{10} \LCI) +  (\log_{10} T )^2 .
}
The relationship between \ERS\ for the data set using time units $T=1$ and $T=1000$ is plotted in Figure \ref{Fig:ERST}.    We see that a simple change of time measurement units essentially scrambles the \ERS\ scores.
\begin{figure}[h]
\centerline{\includegraphics [width= \columnwidth] {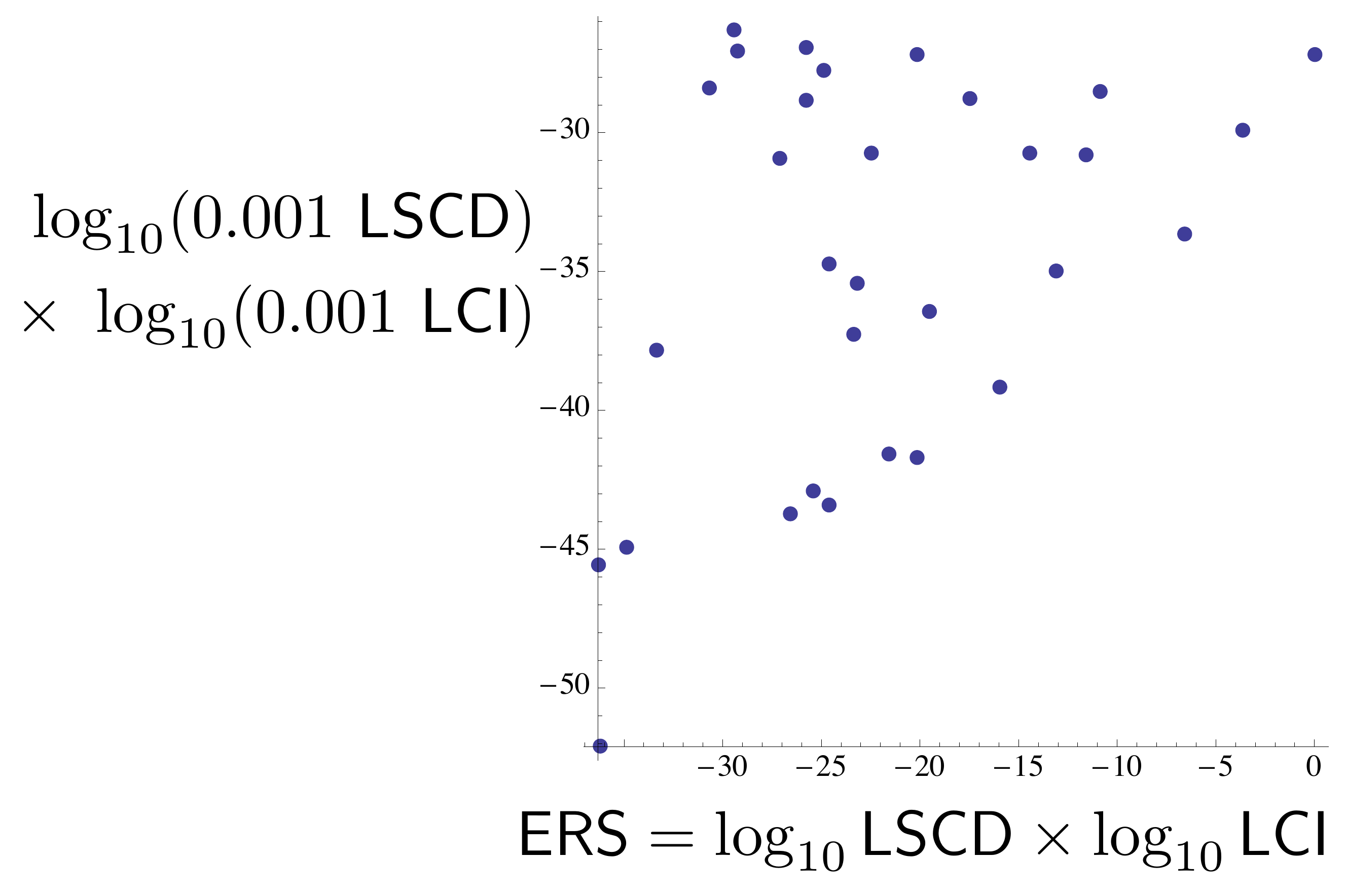}}
\caption{\label {Fig:ERST} \small Non-monotonicity of the ``Extra Risk Score'' (\ERS) under change of time units for the cancer incidence data and the number of stem cell divisions.  Here the time unit is rescaled by a factor of $1/T=0.001$. 
}
\end{figure}

The non-monotonicity of the \ERS\ statistic in \citet{\TV} under a simple change of time units renders any conclusions based on the \ERS\ meaningless.  This includes their Fig. 2, their categorization of ``R-tumors'' and ``D-tumors'', their K-means cluster analysis (Supplement Fig. S2), and therefore the conclusion of their paper (repeated from above):
\Quote{The maximum fraction of
tumors that are preventable through primary
prevention (such as vaccines against infectious
agents or altered lifestyles) may be evaluated
from their ERS. For nonhereditary D-tumors,
this fraction is high and primary prevention could
make a major impact \ldots \hide{Secondary prevention,
obtainable in principle through early detection,
could further reduce nonhereditary D-tumor-
related deaths and is also instrumental for reducing
hereditary D-tumor-related deaths.} For
R-tumors, primary prevention measures are not
likely to be very effective, and secondary prevention
should be the major focus.}

\subsection*{Residual Risk}

Clearly, \citeauthor{\TV}'s $\ERS=\log_{10} \LCI \times \log_{10} \LSCD$ is so ill-behaved that it has no meaning.  The two natural choices for measuring extra risk are:
\enumlist{
\item {\bf Normalized Incidence}:  This is simply the lifetime cancer incidence divided by the lifetime number of stem cell divisions.  In $\log$ scale:  
\ab{
\NI \eqdef \log_{10} \LCI - \log_{10} \LSCD.
}
\item {\bf Residual Risk}:  This is the lifetime cancer incidence divided by the incidence predicted from the linear regression.  In $\log$ scale:  
\ab{
\RR &\eqdef \log_{10} \LCI - \Psf[\log_{10} \LCI ],
}
}	
where
\an{
\Psf[\log_{10} \LCI ] &=  0.533  \log_{10} \LSCD-7.61 \label{eq:Fit}
}
is the linear predictor function from regression of $\log_{10} \LCI$ on $\log_{10} \LSCD$ (plotted with the data in Figure \ref{Fig:Fit}):
\begin{figure}[!h]
\centerline{\includegraphics [width=  \columnwidth] {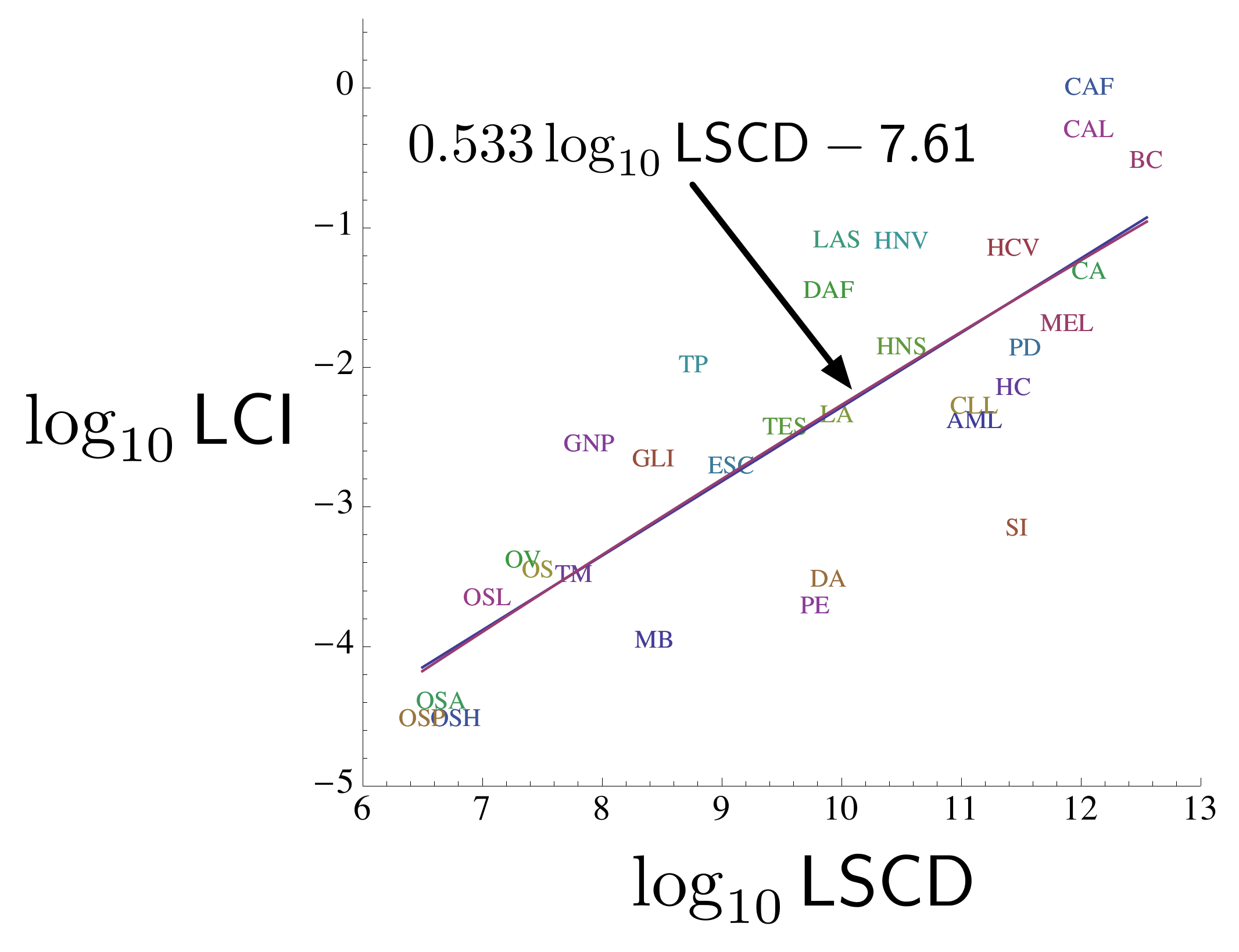}}
\caption{\label {Fig:Fit} The 31 tissues types (abbreviations) with the linear prediction function.}
\end{figure}

In Table \ref{Table:Excess} the 31 tumor types are sorted by their residual risk values, 
\ab{
\RR &= \log_{10} \LCI - (0.533  \log_{10} \LSCD - 7.61).
}
(this has been posted on at least one online blog, \citet{Peer3:2015:1:7}).

\begin{table} [!h]
\caption{\label{Table:Excess} Excess of $\log_{10} \LCI$ above the best fit linear regression: $\RR = \log_{10} \LCI - (0.533  \log_{10} \LSCD-7.61)$.  In {\bf bold face} are the types called ``D-tumors'' in \citet{\TV}.  Their position near the top of the list is due to the coincidentally high rank correlation between $\log_{10} \LCI \times  \log_{10} \LSCD$ and $\log_{10} \LCI - (0.533  \log_{10} \LSCD-7.61)$ on the data set.}
\ab{\footnotesize
\begin{array}{rl}
{\bf \RR} & \text{\bf Tumor Type}\\
\hline
 1.211 & \text{\bf Lung adenocarcinoma (smokers)} \\
 1.183 & \text{\bf Colorectal adenocarcinoma with FAP} \\
 0.952 & \text{\bf Thyroid papillary/follicular carcinoma}
   \\
 0.916 & \text{\bf Head $\&$ neck squamous cell carcinoma
   with HPV-16} \\
 0.886 & \text{\bf Duodenum adenocarcinoma with FAP} \\
 0.882 & \text{\bf Colorectal adenocarcinoma with Lynch
   syndrome} \\
 0.853 & \text{Gallbladder non papillary adenocarcinoma}
   \\
 0.460 & \text{Glioblastoma} \\
 0.403 & \text{\bf Basal cell carcinoma} \\
 0.373 & \text{\bf Hepatocellular carcinoma with HCV} \\
 0.314 & \text{Ovarian germ cell} \\
 0.201 & \text{Osteosarcoma of the legs} \\
 0.178 & \text{Osteosarcoma} \\
 0.156 & \text{Head $\&$ neck squamous cell carcinoma} \\
 0.106 & \text{Testicular germ cell cancer} \\
 0.062 & \text{Esophageal squamous cell carcinoma} \\
 -0.016 & \text{Thyroid medullary carcinoma} \\
 -0.045 & \text{Lung adenocarcinoma (nonsmokers)} \\
 -0.136 & \text{\bf Colorectal adenocarcinoma} \\
 -0.334 & \text{Osteosarcoma of the arms} \\
 -0.373 & \text{Osteosarcoma of the pelvis} \\
 -0.400 & \text{Pancreatic ductal adenocarcinoma} \\
 -0.411 & \text{Melanoma} \\
 -0.520 & \text{Osteosarcoma of the head} \\
 -0.593 & \text{Chronic lymphocytic leukemia} \\
 -0.627 & \text{Hepatocellular carcinoma} \\
 -0.696 & \text{Acute myeloid leukemia} \\
 -0.840 & \text{Medulloblastoma} \\
 -1.181 & \text{Duodenum adenocarcinoma} \\
 -1.312 & \text{Pancreatic endocrine (islet cell)
   carcinoma} \\
 -1.651 & \text{Small intestine adenocarcinoma} \\
 \hline
 0.000 & \text{\bf Total}
\end{array}
}
\end{table}

Topping the list is lung cancer in smokers, followed principally by the tumors labeled with inherited or viral risk factors.  This supports the argument of \cite{\TV} for ``incorporation of a replicative component as a third, quantitative determinant of cancer risk'' in addition ``environmental
or inherited factors.''

We also see something remarkable.  The types identified by \citeauthor{\TV} as ``excess risk'' based on the spurious \ERS\ statistic largely maintain their positions in the top of the list based on the \RR\ statistic.  The reason for this is pure coincidence:  it happens that, on this data set, the correlation between \RR\ and \ERS\ is fortuitously high: $r(\ERS,\RR) = 0.83$, $\rho (\ERS,\RR) = 0.80$.
 
That this is fortuitous can be seen in the completely different mathematical structures of \ERS\ and \RR:  
\ab{
\ERS &=\log_{10} \LCI \times  \log_{10} \LSCD, \\
\RR &= \log_{10} \LCI - (0.533  \log_{10} \LSCD-7.61).
}
For comparison, the rank correlation between \ERS\ and \ERS\ with time units rescaled by $0.001$ is only $0.24$ (Figure \ref{Fig:ERST}), and between \ERS\ and \NI\ it is only $0.09$. Note that $\rho(\NI, \RR) = 0.618$.   

This coincidental correlation may have been the reason that \ERS\ was retained by the authors, because it sorted the tumor types in an order similar to the vertical position of the points relative to the predictor line in Figure \ref {Fig:Fit}.

If one wanted to distinguish ``R-tumors'' from ``D-tumors'' based on Table \ref {Table:Excess}, one certainly could, since there are several large gaps between \RR\ values, notably between Gallbladder and Glioblastoma.  But an goodness-of-fit test \citep{Anderson:and:Darling:1952:Asymptotic} shows the distribution of \RR\ values to be indistinguishable from a Gaussian normal distribution ($P=0.977$), and the gap sizes not significantly different from an exponential distribution ($P=0.51$).  Caution should therefore be used in making any claims based on these gaps.

We also see in Table \ref{Table:Excess} that the residual risk spans $2.862=1.211 + 1.651$ orders of magnitude.   A correlation of $0.8$ between the logarithms of \LSCD\ and \LCI\ therefore does not preclude three orders-of-magnitude variation in \LCI\ due to other factors.  This is elaborated upon in the next section.  If we consider only those tumor types with negative \RR, there are still almost $2$ orders of magnitude variation due to unknown factors.  If not from errors in the data, something is suppressing the incidence of certain tissue tumors to only 2\% of that predicted from the $\log$ \LSCD\ regression.

\section*{The fallacy that \emph{high correlations preclude intervention}}
\label{Section:Robustness}

\citeauthor{\TV} point out that the correlation between $\log_{10} \LSCD$ and $\log_{10} \LCI$ is extremely robust to noise, and therefore makes their results robust.  What they do not realize is that this very robustness argues against their conclusion that ``for R-tumors, primary prevention measures are not likely to be very effective'', because it allows large variation in unknown factors that could also control cancer incidence rates while having little effect on the correlation.

\citeauthor{\TV} measure this robustness by adding noise to the data to see how much it changes the correlation.  They add both Gaussian and uniformly distributed random noise to their estimates of lifetime number of stem cell divisions.  For the uniform variation, they examine the correlations between \LCI\ and  $\LSCD + \Uc (-2,2) $, where $\Uc(-2,2)$ is a uniformly distributed random variable on the interval $[-2,2]$.  Under 10,000 replicates they find the addition of this four orders-of-magnitude noise only reduces the median value of Spearman's rank correlation $\rho$ \citep{Spearman:1904:Proof} from $0.81=\rho(\LCI, \LSCD )$ to $0.67=\rho(\LCI, \LSCD + \Uc (-2,2) )$, which is still significantly different from zero:
\Quote{Thus, though the total range for \LSCD\ is $\sim$ 6 orders of magnitude and we allowed four 4 [sic] orders of magnitude variation for each data point, the correlations generated were always statistically significant.  This provides strong evidence that our results are robust. [Supplement p. 11]
}

But the robustness of the high correlation to order-of-magnitude variations in the data has its converse implication:  it means that high correlation cannot rule out order-of-magnitude variation in other factors that may determine the rates of cancer.  We see this in Table \ref{Table:Excess}. 

To illustrate this converse implication, suppose that \LCI\ were determined by two factors, \LSCD, and another, unknown preventative factor ``\Xf'' that reduces the rate of cancer in proportion to its value.  How much variation in \Xf\ could there be and still obtain correlations of $0.8$ between $\log_{10} \LSCD$ and $\log_{10}\LCI$?

The general problem is to compute the correlation, $r$ \citep{Pearson:1901:LIII}, between $Y$, and $Y$ plus uncorrelated noise, $Z$:
\ab{
r(Y, Y+Z) &= \frac{\Cov(Y, Y+Z)}{\sqrt{\Var(Y)} \sqrt{\Var(Y+Z)}} \\&
=\frac{1}{\sqrt{1+\dspfrac{\Var(Z)}{\Var(Y)}}}.
}
The solution of $r(Y,Y+Z)=0.8$ is $\Var(Z) = 0.5625 \ \Var(Y)$.

To be concrete, let $\LCI = c * \LSCD * \Xf$, where the constant $c$ does not enter into the correlation.  Then 
$\log_{10} \LCI = \log_{10} c + \log_{10} \LSCD + \log_{10} \Xf$.  We let $\log_{10} \LSCD$ be distributed as a uniform random variable on the interval $[{6.50, 12.55}]$, the actual range from the data.  We let $\log_{10} \Xf$ be distributed as an independent random variable uniform on $[-w, 0]$, and ask how big $w$ can be and yet maintain
$0.8 = r(\log_{10} \LCI, \log_{10} \LSCD)$.  The solution is $w=4.5$---four orders of magnitude.

\begin{figure}[!h]
\centerline{\includegraphics [width= \columnwidth] {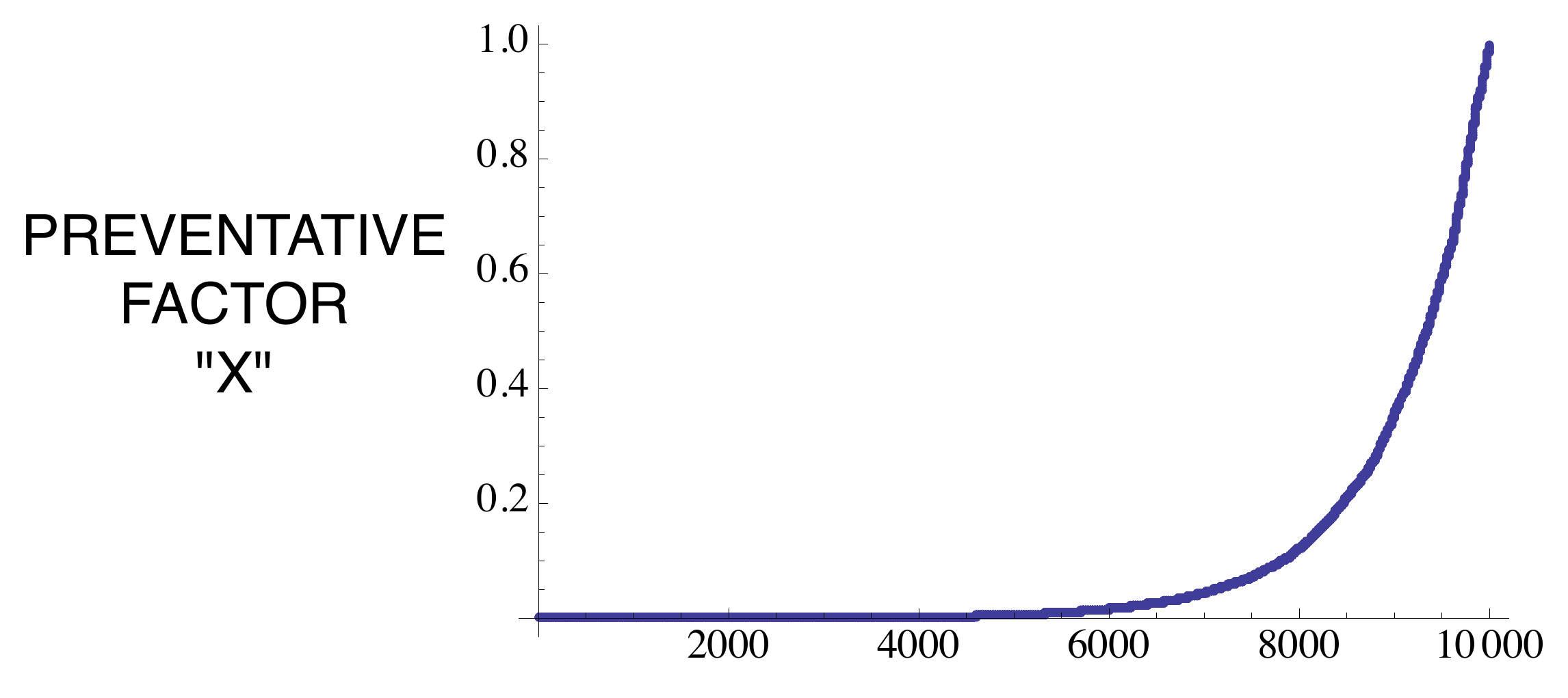}}
\caption{\label {fig:X} Variation in a hypothetical multiplicative ``preventative factor \Xf'' on cancer incidence consistent with a correlation of $0.8$ between $\log_{10} \LSCD$ and $\log_{10} \LCI$.  Sorted along the X-axis are 10,000 samples of \Xf, where $\log_{10} \Xf \sim U[-4.5, 0]$.}
\end{figure}

It may be helpful to see what the distribution of this hypothetical preventative factor $\Xf$ looks like, in Figure \ref{fig:X}.  We see huge variation in the population incidence of cancer due to hypothetical factor $\Xf$, even though there is still a correlation of $0.8$ between $\log_{10} \LSCD$ and $\log_{10} \LCI$.  Therefore, the correlation of $0.8$ does not rule out the presence of other factors---genetic, environmental, physiological, or measurement error---that can produce over four orders-of-magnitude variation in cancer rates.  Such factors could potentially include environmental variation that is subject to intervention.

\section*{Further Explorations}

The problems described above do not diminish the fact that \citet{\TV} have produced an important data set, and the strong correlation they find between the logarithms of lifetime number of stem cell divisions and the rates of cancer in different tissues provides evidence for a real biological phenomenon.  Their data therefore deserve deeper investigation. 

\subsection*{A One-Half Power Law?}

The predictor function \eqref{Fig:Fit} translates to an approximate power law of
\an{
\LCI &\approx 10^{\dsty \, -7.61} \  \LSCD^{\dsty \, 0.533}.\label{eq:PowerLaw}
}

A power law with exponent $0.533$ is not what we would have expected from a simple stem cell division hypothesis.  The simple hypothesis would predict the cancer incidence should grow \emph{in proportion} to the number of stem cell divisions, with exponent $1.00$.  What we find instead is that it grows in proportion to approximately the \emph{square root} of the number of stem cell divisions (this has been noted in comments on at least one online blog, \citet{Kuenzel:2015:1:3}).  
 
The value $0.533$ may, however, reflect \emph{regression dilution}---a systematic reduction in the slope of the predictor function due to large amounts of noise in the {independent} variable.  This is likely here because the estimate of the number of stem cell divisions, \LSCD, combines multiple and sometimes uncertain primary data. Improvements in the estimation of \LSCD\ and inclusion of more tissue types in the data would therefore be expected to increase the slope.  

If, however, further data lent support to a power law with an exponent of one-half, it would invite biological speculation.  One would want to examine probabilistic models for the mutational basis of carcinogenesis to see where probabilities scaled in proportion to the square root of the number of cell divisions.  Random walks and diffusions provide another possibility, since the mean squared displacement grows in proportion to time or the number of steps taken.  In such cases we would need to ask what properties of cells follow a random walk under cell division.  Models of telomere length come to mind here \citep{Blythe:and:MacPhee:2013,Duc:and:Holcman:2013:Computing}.  Another source could be geometry---boundaries of regions that grow along two-dimensions (e.g. the perimeter of a growing circle, or the surface of an elongated cylinder growing in diameter) can have close to a square-root relationship to that growth.  Provided the right geometry, a square-root relationship could conceivably emerge if the boundaries of tissues played a role in carcinogenesis.  There may be plausible sources of one-half power laws to be elicited from the work on scaling relationship in biology (c.f.\ \citet{Savage:et:al:2013:Using}).

\subsection*{The Formula for \LSCD}

The appearance of the exponent $0.533$ in the relationship between $\LCI$ and $\LSCD$ prompts us to examine more closely how the estimate of \LSCD\ was made.  Their Supplement explains that the formula for \LSCD\ is:
\an{
\LSCD = s(2+d)-2, \label{eq:LSCD}
}
where
\desclistZero{
\item[$s$] is the total number of stem cells found in a fully developed tissue, and
\item[$d$] is the number of further divisions of each stem cell  in the lifetime of that tissue once the tissue is fully developed, due to normal tissue turnover.
}

Suppose instead of this formula, we explored the space of possible combinations of the primary data $s$ and $d$ (i.e we are pursuing \emph{symbolic regression} \citep{Koza:1990:GeneticUR}).  Could modern data mining software discover the stem cell division hypothesis by finding patterns in the data (c.f. \citet{Schmidt:and:Lipson:2009:Distilling})?  Could we find combinations of $s$ and $d$ with higher correlations to \LCI?

We first explore the effect of changing the additive constants in \eqref{eq:LSCD}.  For Spearman's $\rho$, the constant $-2$ is irrelevant.  We embed $s(2+d)-2$ in the family of formulae, $\psi(c) \eqdef s(c+d)-2$.  For $c \in [0, 20]$, $\rho(\psi(c), \LCI)$ varies only between $0.776$ and $0.810$, with a broad peak for $0.8 \leq c \leq 6.1$.   The robustness of $\rho$ to variation in the constants means the data give no particular validation to their values in the formula for \LSCD\ \eqref{eq:LSCD}.

Since $\log_{10} \LSCD \approx \log_{10} s + \log_{10} d$, hence \eqref {eq:Fit} is approximated by
\ab{
\Psf[\log_{10} \LCI ] & \approx  0.533 \, ( \log_{10} s +  \log_{10} d) - 7.61.
}
Suppose instead of equal weights on $\log_{10} s$ and $\log_{10} d$, we
embed $s(2+d)-2$ in the family of formulae, 
\ab{
\phi(t) &\eqdef s^{\,\dsty  (2-t)}(2+d^{\dsty \, t})-2.
}
The parameter $t$ varies $\phi(t)$ continuously from $\phi(0) = 3 s^2 - 2$, to $\phi(1) = \LSCD$, to $\phi(2) = d^2$.  We find that 
\ab{
\rho(\phi(0), \LCI) &= \rho(s, \LCI) = 0.68,\\ 
\rho(\phi(1), \LCI) &= \rho(\LSCD, \LCI) = 0.81,\\ 
\rho(\phi(2), \LCI) &= \rho(d, \LCI) = 0.60½. 
}
½
\begin{figure}[h]
\centerline{\includegraphics [width=  \columnwidth] {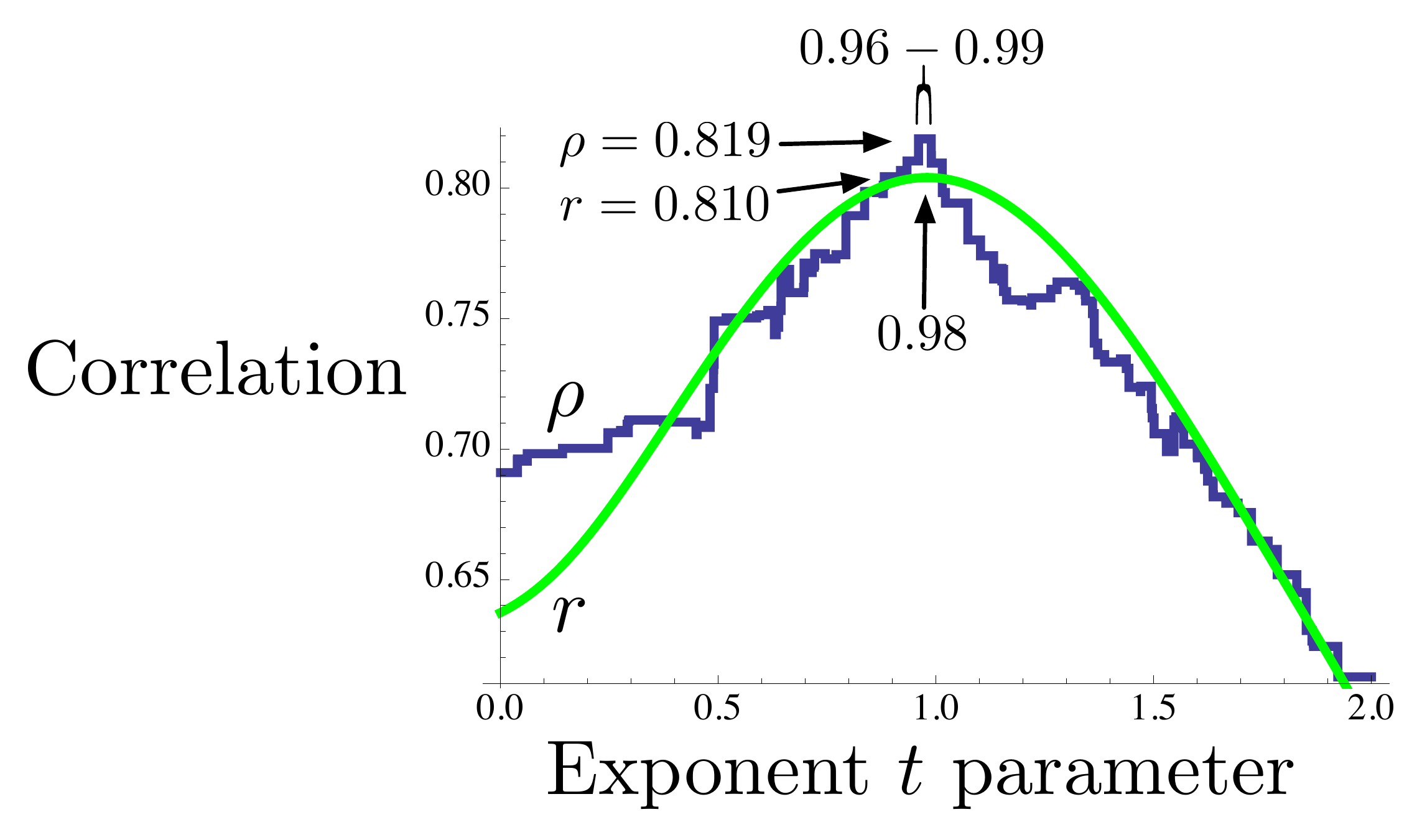}}
\caption{\label {Fig:t} \small Correlations $r(\log_{10} \LCI, \log_{10} \phi(t) )$ and $\rho(\LCI, \phi(t) )$ for $\phi(t) \eqdef s^{\,\dsty  (2-t)}(2+d^{\, \dsty  \,t})-2$.  The peaks are near $t=1$, where $\phi(1) = \LSCD$.}
\end{figure}

For the range of $t$,  Figure \ref{Fig:t}\ plots the correlations $\rho(\phi(t), \LCI)$ and $r(\phi(t), \LCI)$.   There peaks are very near the value $t=1$ where $\phi(1) = \LSCD$.

Therefore, by simply looking for correlations between $\LCI$ and different combinations of $s$ and $d$, we come up with a formula very close to the formula developed by \citeauthor{\TV} \emph{from first principles} for the total number of stem cell divisions in a tissue.  There is no \emph{a priori} reason that the peak should occur near $t=1.0$, since $s$, $d$, and \LCI\ are independently obtained data sets (the rank correlation between $s$ and $d$ is only $0.23$).

Could this be a coincidence?   We can get some idea of that likelihood.  Bootstrap resampling of the 31 tissue types for the values of $t$ that maximize $\rho(\phi(t), \LCI)$ gives a central 50\% interval of $(0.909, 1.009)$, with a median and sharp mode at $t=0.971$.  Subjecting $s$ and $d$ to multiplicative noise (replacing $s$ and $d$ by $s \times \nu_1$ and $d \times \nu_2$, where $\ln \nu_1, \ln \nu_2 \sim \Nc(0,1/8^2)$, $\nu_1, \nu_2$ independent), gives a median of $0.954$ and 50\% central range of $(0.928, 0.973)$.

So, yes it could be a coincidence that the optimal $t$ is so close to $1.0$.  Chance can't be ruled out as to why the formula of \citeauthor{\TV} seems to provide the maximal correlation to the rates of cancer incidence.  We cannot rule out there being other relationships between $s$ and $d$ that give even higher correlations with the cancer incidence data.  It will require expansion of the data set to more tissue types, more precise estimates of the number of stem cell divisions, and exploration of other models of the biology to resolve this question.  

Nonetheless, the optimality of the formula of \citeauthor{\TV} is at least suggestive: that something close to $s \times d$ has a real biological role in the incidence of cancer.  It is a separate form of evidence from the $0.8$ magnitude of the correlation itself.

\section*{Discussion}

The publication of \citet{\TV} was announced by Johns Hopkins University with the press release headline, ``Bad Luck of Random Mutations Plays Predominant Role in Cancer, Study Shows.''   Probability theory was created to make reasoning about ``luck'' a rigorous science.  People's reasoning about probabilities---luck---is a notoriously error-prone activity, and the characterization of these errors is now a scientific field that has burgeoned since the pioneering work of \citep{Tversky:and:Kahneman:1974:Judgment}.  There is a wide intuition that there is a zero-sum tradeoff  between control and luck in determining events:  the more control we have, the less we depend on luck.  This intuition of a tradeoff resonates with the mathematical structure of the analysis of variance, in which total variance is partitioned into a sum of variances and covariances.  However, when two factors, ``luck'' ($L$) and ``control'' ($C$) interact multiplicatively, as $L \times C$, then there is no necessary tradeoff between them at all.  And as described here, because of the nature of logarithms, a high correlation between $\log L$, and $\log L + \log C$, does not preclude large variation in $C$.

Beyond this issue of correlations between logarithms, and the erroneous use of the ill-behaved ``extra risk score'' (\ERS), there is the additional pitfall of cognitive framing.  If \citeauthor{\TV} had presented an 80\% correlation between logarithms of the number of stem cell divisions and cancer \emph{mortality} for different tissues, instead of \emph{incidence}, would they have then argued ``cancer treatment measures are not likely to be very effective, and prevention should be the major focus''?  Of course not, because novel cancer treatments are something new under the sun.  The cognitive frame in this case draws on our familiarity with the history of medicine in which many new cures have been found that obliterate past correlations. 

The experience of ``dramatic cure'' has no parallel in experiences of ``dramatic prevention'' (except to a statistician) because prevention is a non-event.  Therefore, the fallacy that ``high correlation precludes intervention'' can find an easier home when reasoning about prevention.

Cures, primary prevention, and secondary prevention are simply interventions at different stages of disease.  Correlations found in the present do not bear on what interventions may be found in the future for any of these disease stages.  If anything, the track record of human discovery to date has been that cancer prevention is much easier to discover than cancer cure.  But the future has yet to be written.

Leaving aside the erroneous statistics of \citet{\TV} and the conclusions based on them, as  \emph{biology} it is a significant finding that there is a high rank correlation between (1) cancer incidence, and (2) the estimated number of stem cell divisions in a tissue.  Additional upport for that significance is presented here: that among various possible combinations of the primary data, the one that produces the highest correlation to cancer incidence is precisely the formula \citeauthor {\TV} obtain from biological first principles.  Although this could be a statistical coincidence and requires additional data to confirm, it 
can be seen as a novel form of support for the hypothesis stem cell divisions are an independent causal factor for cancer.

{\section*{\small Materials and Methods}
\small Data sets were obtained from the Supplement to \citet{\TV}.  All computations and statistics were carried out using \emph{Mathematica}{\texttrademark}.  The code used is available upon request.
}

\section*{\centerline{\large \sc Acknowledgements}}
I thank my colleagues at the KLI for their spirited discussion---Gerd M\"uller, Isabella Sarto-Jackson, Olivier Morin, and Mathieu Charbonneau, and thank Claus Vogl and Marcus W. Feldman for their comments.   I gratefully acknowledge support from The KLI Institute, Klosterneuburg, Austria, and the Mathematical Biosciences Institute at Ohio State University, USA, through National Science Foundation Award \#DMS 0931642.

{\small

}
\manuscript{\end{spacing}}	
\end{document}